\documentclass[reprint,amsmath,amssymb,superscriptaddress]{revtex4-1}

\usepackage{graphicx}
\usepackage{dcolumn}
\usepackage{bm}
\usepackage{color}

\begin{document}


\title{Theory for tunnel magnetoresistance oscillation}


\author{Keisuke Masuda}
\email{MASUDA.Keisuke@nims.go.jp}
\affiliation{Research Center for Magnetic and Spintronic Materials, National Institute for Materials Science (NIMS), Tsukuba 305-0047, Japan}
\author{Thomas Scheike}
\affiliation{Research Center for Magnetic and Spintronic Materials, National Institute for Materials Science (NIMS), Tsukuba 305-0047, Japan}
\author{Hiroaki Sukegawa}
\affiliation{Research Center for Magnetic and Spintronic Materials, National Institute for Materials Science (NIMS), Tsukuba 305-0047, Japan}
\author{Yusuke Kozuka}
\affiliation{Research Center for Materials Nanoarchitectonics, National Institute for Materials Science (NIMS), Tsukuba 305-0047, Japan}
\author{Seiji Mitani}
\affiliation{Research Center for Magnetic and Spintronic Materials, National Institute for Materials Science (NIMS), Tsukuba 305-0047, Japan}
\affiliation{Graduate School of Science and Technology, University of Tsukuba, Tsukuba 305-8577, Japan}
\author{Yoshio Miura}
\affiliation{Research Center for Magnetic and Spintronic Materials, National Institute for Materials Science (NIMS), Tsukuba 305-0047, Japan}
\affiliation{Faculty of Electrical Engineering and Electronics, Kyoto Institute of Technology, Matsugasaki, Sakyo-ku, Kyoto, 606-8585, Japan}
\affiliation{Center for Spintronics Research Network, Graduate School of Engineering Science, Osaka University, Toyonaka, Osaka 560-8531, Japan}


\date{\today}

\begin{abstract}
The universal oscillation of the tunnel magnetoresistance (TMR) ratio as a function of the insulating barrier thickness in crystalline magnetic tunnel junctions (MTJs) is a long-standing unsolved problem in condensed matter physics. To explain this, we here introduce a superposition of wave functions with opposite spins and different Fermi momenta, based on the fact that spin-flip scattering near the interface provides a hybridization between majority- and minority-spin states. In a typical Fe/MgO/Fe MTJ, we solve the tunneling problem and show that the TMR ratio oscillates with a period of $\sim$\,3\,{\AA} by varying the MgO thickness, consistent with previous and present experimental observations.
\end{abstract}

\pacs{}

\maketitle

The tunneling effect is one of the most fundamental phenomena in quantum mechanics originating from the wave nature of matter. In particular, quantum tunneling has played an important role for various topics in condensed matter physics. For example, in the case of $p$-$n$ junctions, electrons tunnel through the depletion layer under large electric field, giving rise to negative differential resistance \cite{1958Esaki-PR}. As another example, tunneling spectrum in a metal/insulator/superconductor junction provides a clear signature of a gap structure in the density of states of the superconductor, validating the Bardeen-Cooper-Schrieffer theory \cite{1960Giaever-PRL}. Moreover, the scanning tunneling microscope utilizes tunneling electrons for imaging surfaces in the atomic level \cite{1987Binnig-RMP}.

\begin{figure}
\includegraphics[width=8.5cm]{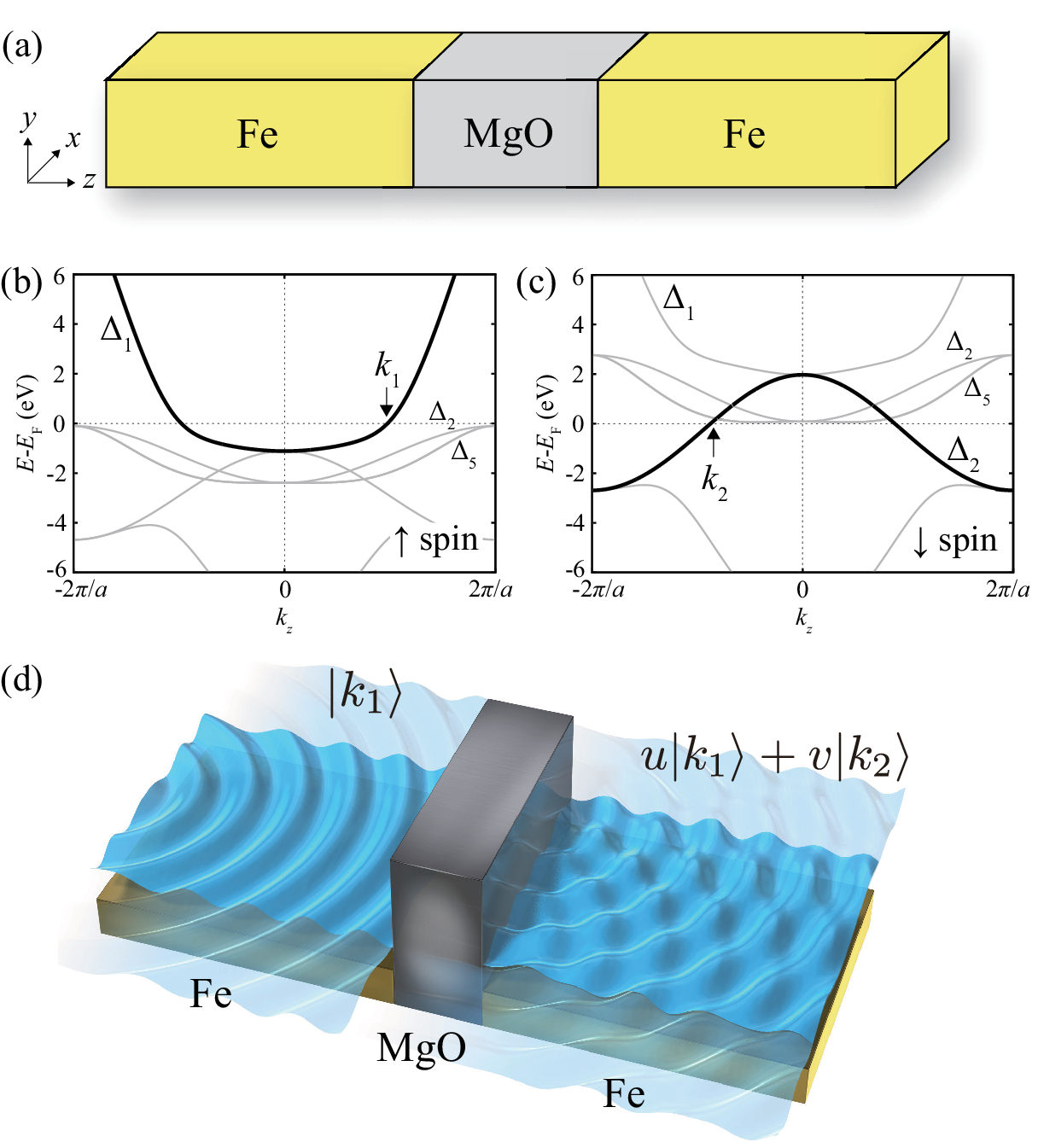}
\caption{\label{Fig1} (a) Schematic of the Fe/MgO/Fe MTJ. (b), (c) Majority ($\uparrow$) -spin and minority ($\downarrow$) -spin band structures of bcc Fe along the $\Delta$ line with ${\bf k}_\parallel=0$. (d) Illustration of our idea including a superposition of wave functions with different Fermi momenta.}
\end{figure}

The tunnel magnetoresistance (TMR) effect is another topic related to tunneling in the field of spintronics. This occurs in magnetic tunnel junctions (MTJs) consisting of an insulating barrier sandwiched between ferromagnetic electrodes [Fig. \ref{Fig1}(a)]. The wave functions in different spin channels have different transmission probabilities because of imbalanced band structures, leading to finite magnetoresistance. One can estimate the magnitude of the magnetoresistance by defining the TMR ratio as a ratio of resistances between parallel and antiparallel magnetization states of the two ferromagnetic electrodes. In 2004, Parkin {\it et al.} \cite{2004Parkin-NatMat} and Yuasa {\it et al.} \cite{2004Yuasa-NatMat} reported significantly high TMR ratios in Fe(Co)/MgO/Fe(Co)(001) MTJs, which provided a basis for further fundamental studies of the TMR effect and their device applications.

However, there is a missing piece in the mechanism of such a giant TMR effect; the universal oscillation of the TMR ratio as a function of the insulating barrier thickness \cite{2004Yuasa-NatMat,2007Matsumoto-APL,2008Ishikawa-JAP,2010Marukame-PRB} has not been explained satisfactorily. In a report of the giant TMR effect in Fe/MgO/Fe(001) \cite{2004Yuasa-NatMat}, Yuasa {\it et al.} observed an oscillation of the TMR ratio with a period of $\sim$\,3\,{\AA} by varying the MgO thickness, referred to as the TMR oscillation. Subsequent experiments \cite{2007Matsumoto-APL} clarified that the TMR oscillation originates from resistance oscillations in both parallel and antiparallel magnetization states. Here, electron tunneling through MgO occurs between the same (different) spin states of the two electrodes in the parallel (antiparallel) magnetization state. Recent experiments for a series of MTJs with high crystallinity \cite{2021Scheike-APL,2022Scheike-APL,2023Scheike-APL} found that the TMR oscillation with a period of $\sim$\,3\,{\AA} is universally observed and its amplitude is much larger than ever reported. Therefore, to elucidate the origin of the TMR oscillation will advance our understanding not only on the TMR effect but also on the quantum tunneling itself. This will also provide guiding principles for achieving even higher TMR ratios.

Conventionally, high TMR ratios in Fe/MgO/Fe(001) have been explained by the $\Delta_1$ coherent tunneling mechanism; the half-metallic $\Delta_1$ band structure of Fe [Figs. \ref{Fig1}(b) and \ref{Fig1}(c)] and the slowest decaying $\Delta_1$ evanescent state of MgO enable a selective tunneling of the perfectly spin-polarized $\Delta_1$ state, leading to a high TMR ratio \cite{2001Butler-PRB,2001Mathon-PRB}. However, the TMR oscillation cannot be explained by this mechanism \cite{2001Butler-PRB,2001Mathon-PRB,2008Heiliger-PRB}. Although additional effects, such as interference of evanescent states \cite{2001Butler-PRB} and nonspecular tunneling \cite{2008Zhang-PRB}, have been considered, these provide a resistance oscillation only in the antiparallel magnetization state, qualitatively in disagreement with the experimental results. Another study \cite{2011Autes-PRB} proposed an oscillation of the TMR ratio due to the quantization in the ferromagnetic layer, but this occurs when varying the thickness of the ferromagnetic layer, inconsistent with the experimental situation.

In this Letter, we show that the TMR oscillation can be explained by taking into account a superposition of wave functions with opposite spins and different Fermi momenta for the tunneling problem. It is known that spin-flip scattering occurs near interfaces of MTJs \cite{2005Mavropoulos-PRB,2011Miura-PRB,2021Masuda-PRBL}, indicating that spin is not a good quantum number in this system. This provides a hybridization between majority- and minority-spin states with different Fermi momenta [Figs. \ref{Fig1}(b) and \ref{Fig1}(c)], which justifies our assumption on the superposition of wave functions. Focusing on Fe/MgO/Fe(001), we solve tunneling problems assuming a superposition of majority-spin $\Delta_1$ and minority-spin $\Delta_2$ wave functions with different Fermi momenta as a transmitted wave function [see Fig. \ref{Fig1}(d)]. We obtain transmittances in the parallel and antiparallel magnetization states, from which the TMR ratio is calculated. It is found that the transmittances and the TMR ratio have oscillatory behaviors with a period of $\sim$\,3\,{\AA} as a function of the MgO thickness, in agreement with previous and present experimental observations. We also show that the calculated TMR ratio can reproduce our experimental results not only qualitatively but also quantitatively by tuning the parameters in our model. Although we focus on the TMR oscillation in this Letter, the superposition of wave functions with different Fermi momenta is a general concept and would be helpful to understand transport properties in other tunnel junctions with superconductors, semiconductors, etc.

To make the point of our approach clearer, we start by reviewing the conventional analytical treatment of the tunneling problem in an MTJ. Let us consider the situation that the wave function in the left electrode propagates to the right electrode passing through the insulating barrier, which is described by a coordinate system with the $z$ axis along the stacking direction of the MTJ [Fig. \ref{Fig1}(a)]. For simplicity, we focus on the wave functions with ${\bf k}_\parallel=(k_x,k_y)=(0,0)$ providing the dominant contribution to tunneling transport. When the Fermi momentum is given by $k_z=k_{\rm L}$ ($k_{\rm R}$) in the left (right) electrode, the wave function $\psi_{\rm L}$ ($\psi_{\rm R}$) in the left (right) electrode and the wave function $\psi_{\rm b}$ in the insulating barrier are expressed as
\begin{eqnarray}
\psi_{\rm L}(z)&=&e^{ik_{\rm L}z}+R\,e^{-ik_{\rm L}z},\\
\psi_{\rm b}(z)&=&A\,e^{-\kappa z}+B\,e^{\kappa z},\\
\psi_{\rm R}(z)&=&C\,e^{ik_{\rm R}z}, \label{conv_tun_prob}
\end{eqnarray}
where $\kappa$ is the decaying wave number inside the insulating barrier. After determining $R$, $A$, $B$, and $C$ from continuation conditions for the wave function and its derivative at $z=0$ and $d$ (see the Appendix \ref{Appe_B} for details), we find the following expression for the transmittance:
\begin{eqnarray}
\nonumber T\!=\!\frac{ 16\,\tilde{k}_{\rm L} \kappa^2 \tilde{k}_{\rm R} e^{2\kappa d} }{ \left[\kappa (\tilde{k}_{\rm L}+\tilde{k}_{\rm R}) \left(1+e^{2\kappa d} \right)\right]^2\!\!\!+\!\left[(\kappa^2-\tilde{k}_{\rm L} \tilde{k}_{\rm R})(1-e^{2\kappa d})\right]^2 },\\
\label{conv_trans}
\end{eqnarray}
where $d$ is the thickness of the insulating barrier, $\tilde{k}_{\rm L}=(m_{\rm b}/m_{\rm L})\,k_{\rm L}$, and $\tilde{k}_{\rm R}=(m_{\rm b}/m_{\rm R})\,k_{\rm R}$. Here, $m_{\rm L(R)}$ and $m_{\rm b}$ are the effective masses in the left (right) electrode and the insulating barrier, respectively \cite{remark_conv_trans}. We can obtain the conductance $G$ by substituting Eq. (\ref{conv_trans}) into the Landauer formula $G=(e^2/h)\,T$. The $\Delta_1$ coherent tunneling mechanism mentioned above can be confirmed by employing this tunneling theory in combination with the first-principles calculation \cite{2012Miura-PRB,2017Masuda-JJAP,2017Masuda-PRB}.

However, this conventional tunneling theory cannot describe the oscillation of the TMR ratio in the Fe/MgO/Fe(001) MTJ. Actually, the transmittance in the parallel magnetization state is obtained by putting $k_{\rm L}=k_{\rm R}=k_1$ (or $k_2$) in Eq. (\ref{conv_trans}), where $k_1$ and $k_2$ are the Fermi momenta of the majority-spin $\Delta_1$ and the minority-spin $\Delta_2$ bands of Fe, respectively [see Figs. \ref{Fig1}(b) and \ref{Fig1}(c)]. Note that the negative $k_2$ value with a positive group velocity is chosen for the minority-spin state, since we focus on right-moving states. The transmittance in the antiparallel magnetization state is similarly obtained by setting $k_{\rm L}=k_1$ (or $k_2$) and $k_{\rm R}=k_2$ (or $k_1$) in Eq. (\ref{conv_trans}). As seen from Eq. (\ref{conv_trans}), both the parallel and antiparallel transmittances decrease exponentially with increasing $d$ in monotonic manner without any oscillations.

To explain the oscillation in the transmittance, we introduce a superposition of wave functions between the majority-spin $\Delta_1$ and minority-spin $\Delta_2$ states for the transmitted wave in the tunneling problem \cite{remark_superposition}. Details on the choice of these wave functions are discussed in the Appendix \ref{Appe_D}. Let us first calculate the parallel transmittance $T_{\rm P}$. Based on the fact that the majority-spin $\Delta_1$ state provides the dominant contribution to the TMR effect \cite{2001Butler-PRB,2001Mathon-PRB}, $T_{\rm P}$ can be calculated as $T_{\rm P}=T_{\rm P,\uparrow}+T_{\rm P,\downarrow} \approx T_{\rm P,\uparrow}$, where $\uparrow$ ($\downarrow$) indicates that tunneling electrons are in the majority-spin (minority-spin) state in the left electrode. For the calculation of $T_{\rm P,\uparrow}$, we consider a tunneling from the majority-spin $\Delta_1$ state in the left electrode to the superposition state in the right electrode with the dominant contribution from the majority-spin $\Delta_1$ state, which is given by
\begin{eqnarray}
\psi_{\rm L}(z)&=&e^{ik_1 z}+R\,e^{-ik_1 z},\\
\psi_{\rm b}(z)&=&A\,e^{-\kappa z}+B\,e^{\kappa z},\\
\psi_{\rm R}(z)&=&C\,(u\,e^{ik_1 z}+v\,e^{ik_2 z}). \label{new_tun_prob}
\end{eqnarray}
Here, $u$ and $v$ are matrix elements of the unitary matrix that diagonalizes the $2 \times 2$ Hamiltonian including the effect of the interfacial spin-flip scattering as off-diagonal elements. An explicit expression of the Hamiltonian and the derivation of the wave function in Eq. (\ref{new_tun_prob}) are given in the Appendix \ref{Appe_E}. We impose $|u|\gg|v|$ because of the dominance of the majority-spin $\Delta_1$ state in $T_{\rm P,\uparrow}$. The matrix elements also satisfy the normalization condition $|u|^2+|v|^2=1$. By using the continuity of the wave function and its derivative at $z=0$ and $d$ \cite{remark_conti-condition}, we can derive the following expression for $T_{{\rm P},\uparrow}$:
\begin{widetext}
\begin{eqnarray}
T_{{\rm P},\uparrow}&&\,=\tilde{k}^{-1}_{1{\rm L}}\left[ \tilde{u}^2 \tilde{k}_{1{\rm R}}+\tilde{v}^2 \tilde{k}_{2{\rm R}}+\tilde{u}\tilde{v}\,(\tilde{k}_{1{\rm R}}+\tilde{k}_{2{\rm R}})\cos{((k_1-k_2)d-\theta)} \right]|C|^2,\label{new_trans}\\
\nonumber {\rm Denominator\,\, of}\,\,|C|^2&&\,=(e^{\kappa d}-e^{-\kappa d})^2 \left\{ \kappa^4\left[1+2\,\tilde{u}\tilde{v}\cos{((k_1-k_2)d-\theta)}\right] \right.\\
\nonumber &&\hspace{2.3cm}-2\,\kappa^2\tilde{k}_{1{\rm L}}\left[\tilde{u}^2\tilde{k}_{1{\rm R}}+\tilde{v}^2\tilde{k}_{2{\rm R}}+\tilde{u}\tilde{v}\,(\tilde{k}_{1{\rm R}}+\tilde{k}_{2{\rm R}})\cos{((k_1-k_2)d-\theta)}\right] \\
\nonumber&&\hspace{2.3cm}\left. +\tilde{k}^2_{1{\rm L}}\left[\tilde{u}^2\tilde{k}^2_{1{\rm R}}+\tilde{v}^2\tilde{k}^2_{2{\rm R}}+2\,\tilde{u}\tilde{v}\,\tilde{k}_{1{\rm R}}\tilde{k}_{2{\rm R}}\cos{((k_1-k_2)d-\theta)}\right] \right\}\\
\nonumber &&\,\,+(e^{\kappa d}+e^{-\kappa d})^2 \left\{ \kappa^2\tilde{k}^2_{1{\rm L}}\left[1+2\,\tilde{u}\tilde{v}\cos{((k_1-k_2)d-\theta)}\right] \right.\\
\nonumber &&\hspace{2.3cm}+2\,\kappa^2\tilde{k}_{1{\rm L}}\left[\tilde{u}^2\tilde{k}_{1{\rm R}}+\tilde{v}^2\tilde{k}_{2{\rm R}}+\tilde{u}\tilde{v}\,(\tilde{k}_{1{\rm R}}+\tilde{k}_{2{\rm R}})\cos{((k_1-k_2)d-\theta)}\right] \\
\nonumber&&\hspace{2.3cm}\left. +\kappa^2\left[\tilde{u}^2\tilde{k}^2_{1{\rm R}}+\tilde{v}^2\tilde{k}^2_{2{\rm R}}+2\,\tilde{u}\tilde{v}\,\tilde{k}_{1{\rm R}}\tilde{k}_{2{\rm R}}\cos{((k_1-k_2)d-\theta)}\right] \right\}\label{Deno_C2}\\
&&\,\,+2\,(e^{\kappa d}+e^{-\kappa d})(e^{\kappa d}-e^{-\kappa d})\,\tilde{u}\tilde{v}\,(\kappa^2+\tilde{k}^2_{1{\rm L}})\,\kappa\,(\tilde{k}_{1{\rm R}}-\tilde{k}_{2{\rm R}})\sin{((k_1-k_2)d-\theta)} \\
{\rm Numerator\,\, of}\,\,|C|^2&&=16\,\tilde{k}^2_{1{\rm L}}\,\kappa^2,\label{Nume_C2}
\end{eqnarray}
\end{widetext}
where $\tilde{k}_{1{\rm L}}=(m_{\rm b}/m_{\rm L})\,k_1$, $\tilde{k}_{1{\rm R}}=(m_{\rm b}/m_{\rm R})\,k_1$, and $\tilde{k}_{2{\rm R}}=(m_{\rm b}/m_{\rm R})\,k_2$. We put $u=\tilde{u}$ and $v=\tilde{v}\,e^{i\theta}$ using positive real numbers $\tilde{u}$ and $\tilde{v}$. The relation $\tilde{u}^2+\tilde{v}^2=1$ was used to simplify the expression. Equations (\ref{new_trans}) and (\ref{Deno_C2}) include several terms with $\cos{((k_1-k_2)\,d-\theta)}$ or $\sin{((k_1-k_2)\,d-\theta)}$, leading to an oscillation of the transmittance as a function of $d$. Physically speaking, this oscillation originates from the interference of the majority-spin $\Delta_1$ and the minority-spin $\Delta_2$ wave functions in the transmitted wave, which is seen from the analogy with the double-slit experiment in elementary quantum mechanics (see the Appendix \ref{Appe_F} for details). The antiparallel transmittance $T_{{\rm AP},\uparrow}$ is easily obtained by replacing $\tilde{u}$ with $\tilde{v}$ and $\tilde{v}$ with $-\tilde{u}$ in Eqs. (\ref{new_trans})--(\ref{Nume_C2}) \cite{remark_AP-state}. Since $\tilde{u} \gg \tilde{v}$, this replacement allows us to consider the transmittance for the electron tunneling from the majority-spin $\Delta_1$ state in the left electrode to the superposition state in the right electrode with the dominant contribution from the minority-spin $\Delta_2$ state, which corresponds to $T_{{\rm AP},\uparrow}$. Using $T_{{\rm AP},\uparrow}$, the total antiparallel transmittance can be calculated as $T_{\rm AP}=T_{{\rm AP},\uparrow}+T_{{\rm AP},\downarrow} \approx 2\,T_{{\rm AP},\uparrow}$. We simply set $\tilde{u}=0.95$, $\theta=0$, $m_{\rm b}/m_{\rm L}=1.0$, and $m_{\rm b}/m_{\rm R}=1.0$ (--1.0) for the numerical calculation of $T_{{\rm P},\uparrow}$ ($T_{{\rm AP},\uparrow}$); however, the period and overall shape of the oscillation in the transmittance do not change if we change $\tilde{u}$ within the range of $1>\tilde{u}\gg\tilde{v}$. Note here that $m_{\rm b}/m_{\rm L}$ and $m_{\rm b}/m_{\rm R}$ need to have different signs in the antiparallel state, which is discussed in the Appendix \ref{Appe_C}.

\begin{figure}
\includegraphics[width=8.5cm]{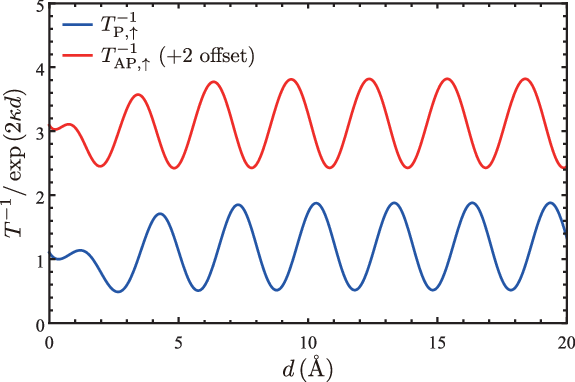}
\caption{\label{Fig2} Barrier thickness $d$ dependencies of inverses of parallel and antiparallel transmittances, $T^{-1}_{{\rm P},\uparrow}$ and $T^{-1}_{{\rm AP},\uparrow}$, divided by $\exp{(2\kappa d)}$.}
\end{figure}
\begin{figure}
\includegraphics[width=8.5cm]{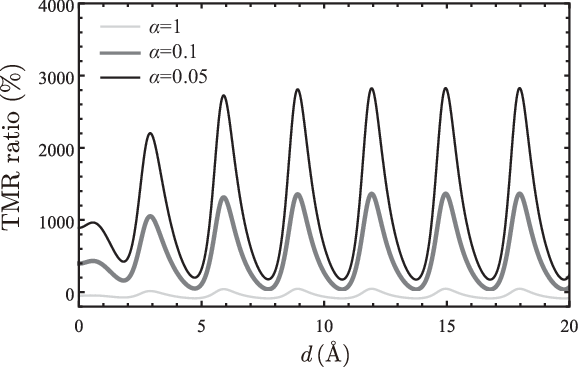}
\caption{\label{Fig3} TMR ratios as a function of the barrier thickness $d$ for different values of $\alpha$. This was obtained by Eq. (\ref{TMR1}).}
\end{figure}

Figure \ref{Fig2} shows inverses of parallel and antiparallel transmittances, $T_{{\rm P},\uparrow}^{-1}$ and $T_{{\rm AP},\uparrow}^{-1}$, divided by the exponentially increasing factor $\exp{(2\kappa d)}$. Here, we set $\kappa=0.2\pi/a_{\rm MgO}$ ($a_{\rm MgO}=4.217\,{\rm \AA}$: lattice constant of MgO), which is the decaying wave number for the $\Delta_1$ complex band of MgO calculated by the {\scriptsize PWCOND} code \cite{2004Smogunov-PRB}. We also used $k_1=1.0\pi/a_{\rm Fe}$ and $k_2=-0.9\pi/a_{\rm Fe}$ ($a_{\rm Fe}=2.866\,{\rm \AA}$: lattice constant of bcc Fe) obtained by calculating the band structure of bcc Fe [Figs. \ref{Fig1}(b) and \ref{Fig1}(c)] with the aid of {\scriptsize QUANTUM ESPRESSO} \cite{2009Giannozzi-JPCM}. In both $T_{{\rm P},\uparrow}^{-1}$ and $T_{{\rm AP},\uparrow}^{-1}$, we can see a clear oscillation with a period of $2\pi/(k_1-k_2)$. From the values of $k_1$ and $k_2$ mentioned above, the period is estimated to be $\sim3\,{\rm \AA}$. These are consistent with the experimental fact that resistances in both the parallel and antiparallel magnetization states have oscillatory barrier thickness dependencies with periods of $\sim3\,{\rm \AA}$ \cite{2007Matsumoto-APL,2022Scheike-APL,2023Scheike-APL}. Finally, we would like to point out that either shape or phase in the oscillation needs to have a difference between $T_{{\rm P},\uparrow}^{-1}$ and $T_{{\rm AP},\uparrow}^{-1}$ for the occurrence of the TMR oscillation shown in Fig. \ref{Fig3} \cite{remark_osci-phase1}.

\begin{figure}
\includegraphics[width=8.5cm]{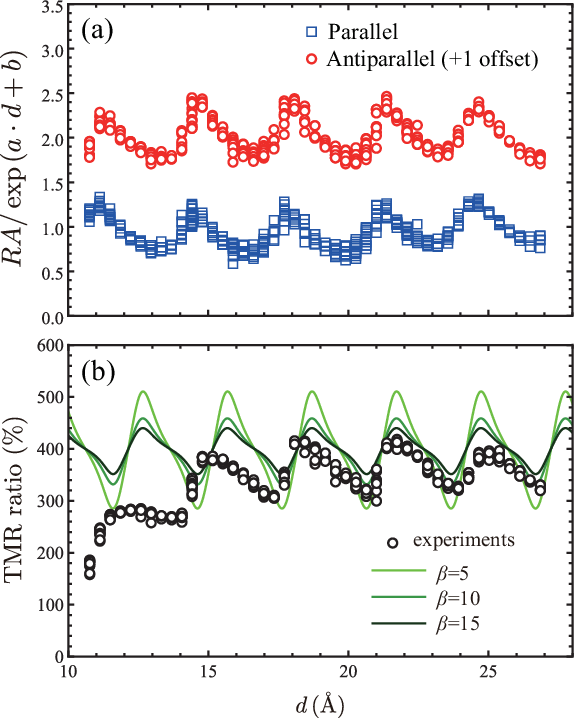}
\caption{\label{Fig4} Experimental results of (a) $RA$ values and (b) TMR ratios in Fe/Mg$_4$AlO$_x$/Fe(001) at room temperature. Panel (b) also shows theoretical values of TMR ratios for different values of $\beta$ calculated by Eq. (\ref{TMR2}) with $\alpha=0.1$ \cite{remark_d-shift}.}
\end{figure}

By using $T_{{\rm P},\uparrow}$ and $T_{{\rm AP},\uparrow}$, we calculated the TMR ratio given by
\begin{equation}
(T_{\rm P}\!-\!\alpha\,T_{\rm AP})/\alpha\,T_{\rm AP} \approx (T_{{\rm P},\uparrow}\!-\!2\,\alpha\,T_{{\rm AP},\uparrow})/2\,\alpha\,T_{{\rm AP,\uparrow}}.\label{TMR1}
\end{equation}
Here, we introduced an electronic-structure parameter $\alpha$, by which $T_{\rm AP}$ is scaled relative to $T_{\rm P}$ reflecting electronic structures of Fe and MgO. The case with $\alpha=1$ corresponds to the usual definition of the TMR ratio. However, as shown in Fig. \ref{Fig3}, the TMR ratio takes negative values for $\alpha=1$, inconsistent with positive high TMR ratios observed in experiments. This is because the present analysis employs only the values of the Fermi momenta and the decaying wave number and does not consider detailed electronic structures of Fe and MgO. If these electronic structures are taken into account by using the first-principles calculation \cite{2001Butler-PRB,2001Mathon-PRB}, values of $T_{\rm AP}$ are around one order of magnitude smaller than those of $T_{\rm P}$. Thus, we set $\alpha=0.1$ for a better comparison with experimental values of the TMR ratio obtained at room temperature in Fig. \ref{Fig4}(b) \cite{remark1}. Figure \ref{Fig3} shows barrier thickness $d$ dependencies of the TMR ratio for different values of $\alpha$. For all values of $\alpha$, the TMR ratio shows an oscillation with a period of $2\pi/(k_1-k_2)\sim3\,{\rm \AA}$, similarly to $T^{-1}_{{\rm P},\uparrow}$ and $T^{-1}_{{\rm AP},\uparrow}$.

Let us directly compare our calculation results with an experimentally observed TMR oscillation. To this aim, we fabricated an MTJ structure and measured magnetotransport properties. The experimental method is explained in the Appendix \ref{Appe_G}. Figure \ref{Fig4}(a) shows barrier thickness $d$ dependencies of the resistance-area product ($RA$) in the parallel and antiparallel magnetization states, where $RA$ is a product of the resistance and the cross-sectional area of the MTJ. We show values of the $RA$ divided by $\exp{(a \cdot d+b)}$, where $a$ and $b$ were determined to be 5.48 (5.73)\,nm$^{-1}$ and $-2.36$ ($-1.35$) in the parallel (antiparallel) magnetization state, respectively, from the fits using the exponential function. As seen in Fig. \ref{Fig4}(a), the $RA$ has an oscillatory $d$ dependence with a period of $3.1\,{\rm \AA}$ in both the parallel and antiparallel magnetization states \cite{remark_osci-amp}. We also find that both the shape and phase of the $RA$ oscillation are slightly different between the parallel and antiparallel magnetization states \cite{remark_osci-phase2}, leading to the TMR oscillation in Fig. \ref{Fig4}(b). For a direct comparison between theoretical and experimental results, we recalculated the TMR ratio using
\begin{equation}
[(T_{\rm P}+\beta\, e^{-2 \kappa d})-\alpha\,(T_{\rm AP}+\beta\, e^{-2 \kappa d})]/\alpha\,(T_{\rm AP}+\beta\, e^{-2 \kappa d}),\label{TMR2}
\end{equation}
where $\alpha$ is fixed to 0.1. Note here that another parameter $\beta$ is introduced for transmittances at ${\bf k}_{\parallel} \neq 0$ \cite{remark2}. In the derivation of Eqs. (\ref{new_trans})--(\ref{Nume_C2}), we considered only the electronic states at ${\bf k}_{\parallel}=0$ because of its dominant contribution to the TMR effect. However, in actual experiments, electronic states at ${\bf k}_{\parallel} \neq 0$ can also contribute to the transmission, which is expressed by the $\beta$-related terms in Eq. (\ref{TMR2}). Since transmittances at different ${\bf k}_{\parallel}$ should have different periods of oscillations, these oscillations are mixed and cancel each other. Thus, we treated the $\beta$-related terms as simple exponential functions. In Fig. \ref{Fig4}(b), calculated TMR ratios for different values of $\beta$ are compared with experimental results. We find that a saw-tooth-like shape of the TMR oscillation is quite similar between theoretical and experimental results \cite{remark3}. In addition, the TMR ratios calculated for $\beta=15$ are found to quantitatively agree with experimental values. Therefore, we conclude that the TMR ratio calculated by Eq. (\ref{TMR2}) can reproduce the experimental results not only qualitatively but also quantitatively. To further validate our theory, we have conducted additional experiments and discussed the relation with previous experiments using Heusler alloys, which are explained in the Appendices \ref{Appe_H} and \ref{Appe_I}.

In summary, we proposed a theory for explaining the universal oscillation of the TMR ratio called the TMR oscillation. Based on the fact that spin-flip scattering occurs near interfaces of MTJs, we took into account the superposition of the majority-spin $\Delta_1$ and minority-spin $\Delta_2$ wave functions with different Fermi momenta for the tunneling problem in Fe/MgO/Fe(001). We analytically calculated transmittances in the parallel and antiparallel magnetization states, from which the TMR ratio was obtained. It was found that the transmittances and the TMR ratio have oscillatory barrier thickness dependencies with a period of $\sim3\,{\rm \AA}$, consistent with the experimental observations. According to our theory, the period of the TMR oscillation is determined by the difference of the Fermi momenta between the majority- and minority-spin states in the ferromagnetic electrode. Therefore, the period of $\sim3\,{\rm \AA}$ is specific to bcc Fe used as electrodes. If MTJs with other ferromagnetic electrodes are successfully made, TMR oscillations with periods different from $3\,{\rm \AA}$ would be observed. We expect future experimental studies using a wider range of materials will provide further information for the TMR oscillation.

\begin{acknowledgments}
The authors are grateful to S. Yuasa and H. Imamura for fruitful discussions. The authors also thank S. Kasai for magnetotransport measurements and H. Ikeda for her technical support on device microfabrication. This work was supported by Grants-in-Aid for Scientific Research (Grants No. 22H04966, No. 23K03933, and No. 24H00408) and MEXT Program: Data Creation and Utilization-Type Material Research and Development Project (Grant No. JPMXP1122715503). MANA is supported by the World Premier International Research Center Initiative (WPI) of MEXT, Japan. The band structure calculation was performed on the Numerical Materials Simulator at NIMS.
\end{acknowledgments}

\appendix
\section{\label{Appe_A} Basis of our theory}
In this study, we discuss the tunnel magnetoresistance (TMR) oscillation by solving the tunneling problem constructed based on the band structure in each region of the Fe/MgO/Fe(001) magnetic tunnel junction (MTJ). We here adopt an effective-mass approximation; an electron in each band at the Fermi level ($E_{\rm F}$) is treated as a free electron with an effective mass and its wave function is expressed as a plane wave. This is equivalent to approximating each band near $E_{\rm F}$ as a parabolic curve; namely, we treat the majority-spin $\Delta_1$ and the minority-spin $\Delta_2$ bands in Fe [thick curves in Figs. 1(b) and 1(c)] as downward-convex and upward-convex parabolas, respectively. Such a simplification allows us to obtain analytical expressions for the transmittance and the period of the oscillation [Eqs. (8)--(10)]. Note that although the effective mass for each band is considered within the parabolic approximation, values of the Fermi momenta used in our study are those obtained by the first-principles calculation. This is because the period of the oscillation $2\pi/(k_1-k_2)$ includes the Fermi momenta ($k_1$ and $k_2$) and their precise values are needed for the direct comparison with experimental results.

\section{\label{Appe_B} Continuation conditions in the tunneling problem}
We use the following continuation conditions for the wave function and its derivative:
\begin{eqnarray}
\psi_{\rm L}(0)&=&\psi_{\rm b}(0),\\
\psi_{\rm b}(d)&=&\psi_{\rm R}(d),\\ 
m^{-1}_{\rm L}\psi^{\prime}_{\rm L}(0)&=&m^{-1}_{\rm b}\psi^{\prime}_{\rm b}(0),\\
m^{-1}_{\rm b}\psi^{\prime}_{\rm b}(d)&=&m^{-1}_{\rm R}\psi^{\prime}_{\rm R}(d),
\end{eqnarray}
where $m_{\rm L(R)}$ and $m_{\rm b}$ are the effective masses in the left (right) electrode and the insulating barrier, respectively. These continuation conditions including effective masses were originally introduced for discussing electron tunneling in semiconductor junctions described by the effective-mass model \cite{1998Davies}. In this model, the total wave function of a semiconductor is expressed by the product of the envelope function and the Bloch function, where the envelope function represents the envelope (i.e., overall shape) of the total wave function. Importantly, continuation conditions for the envelope function and its derivative need to be considered for electron tunneling in semiconductor junctions, as mentioned in Refs. \cite{1998Davies} and \cite{1966BenDaniel-PR}. Therefore, all the wave functions in the present work should be interpreted as the envelope functions in a more precise sense.

\section{\label{Appe_C} Treatment of effective masses}
As we mentioned in the Appendix \ref{Appe_A}, the majority-spin $\Delta_1$ (minority-spin $\Delta_2$) band in Fe is approximated by a downward-convex (upward-convex) parabola in this study. Thus, the approximated $\Delta_1$ and $\Delta_2$ bands have positive and negative values of the effective mass $m^{\ast}$, respectively, following the expression $m^{\ast}=\hbar^2 (d^2E/dk^2_z)^{-1}$. This results in $m_{\rm L}>0$ since the majority-spin $\Delta_1$ band is considered for the incident wave in the left Fe electrode. In the right Fe electrode, we can set $m_{\rm R}>0$ ($m_{\rm R}<0$) in the parallel (antiparallel) magnetization state of the electrodes because the majority-spin $\Delta_1$ (minority-spin $\Delta_2$) state provides the dominant contribution to the superposition state in Eq. (7). We can also set $m_{\rm b}>0$ since the conduction band of the MgO barrier is approximated by a downward-convex parabola with a positive effective mass. Considering these signs of effective masses, we simply put $m_{\rm b}/m_{\rm L}=1$ and $m_{\rm b}/m_{\rm R}=1$ (--1) for the parallel (antiparallel) magnetization state of the electrodes when we numerically calculate the transmittances and the TMR ratio.

\begin{figure}
\includegraphics[width=8.0cm]{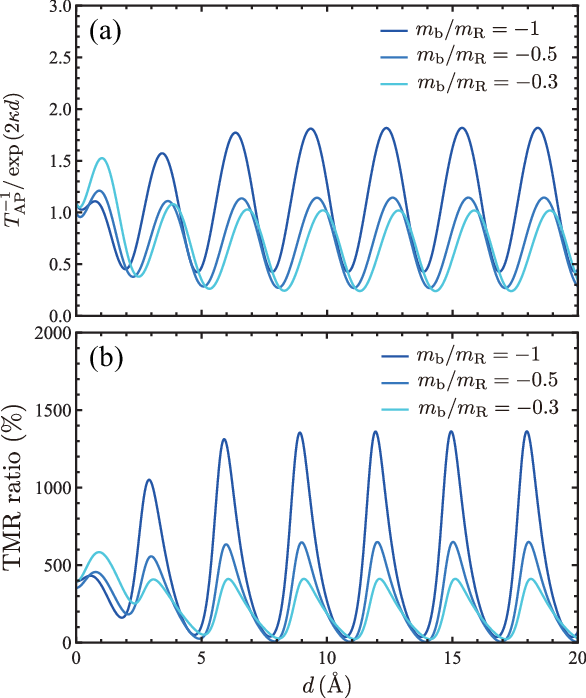}
\caption{\label{FigS1} Barrier thickness $d$ dependences of (a) $T^{-1}_{\rm AP}$ and (b) the TMR ratio calculated using $m_{\rm b}/m_{\rm R}=-1,\,-0.5,\,-0.3$ in the antiparallel magnetization state of the electrodes.}
\end{figure}

However, the absolute value of the effective mass in the electrode is usually different from that in the tunnel barrier. For example, if we consider a more precise energy dispersion of the minority-spin $\Delta_2$ band obtained by the first-principles calculation [Fig. 1(c)], the effective mass of this band should have a large absolute value because of a small value of $|d^2E/dk^2_z|$ around $E_{\rm F}$, leading to a small value of $|m_{\rm b}/m_{\rm R}|$ in the antiparallel magnetization state. Thus, we numerically calculated the inverse of the transmittance $T^{-1}_{\rm AP}$ and the TMR ratio using smaller values of $|m_{\rm b}/m_{\rm R}|$ in the antiparallel magnetization state as shown in Figs. \ref{FigS1}(a) and \ref{FigS1}(b). We see that the overall shape of the oscillation hardly changes for these values of $m_{\rm b}/m_{\rm R}$ in both $T^{-1}_{\rm AP}$ and the TMR ratio. It should also be emphasized that the period of the oscillation does not depend on the values of $m_{\rm b}/m_{\rm R}$ since it is analytically given by $2\pi/(k_1-k_2)$. These results indicate that the qualitative features of the TMR oscillation obtained in this work are insensitive to the choice of parameters for effective masses.

\section{\label{Appe_D} Choice of wave functions}
We consider a superposition of wave functions between the majority-spin $\Delta_1$ and the minority-spin $\Delta_2$ states, characterized by the Fermi momenta $k_1\,(>0)$ and $k_2\,(<0)$, for the transmitted wave in the tunneling problem. There is another minority-spin $\Delta_5$ band near the Fermi level as shown in Fig. 1(c). We easily see that this band has a nearly flat dispersion close to the Fermi level; namely, the group velocity $dE/dk_z$ of this band is quite small. This means that there is a large mismatch in the group velocity between the majority-spin $\Delta_1$ state at $k_z=k_1$ and the above minority-spin $\Delta_5$ state. Therefore, it is unnatural to consider the superposition between these two states. In addition, even if there is a contribution from the minority-spin $\Delta_5$ state to the TMR oscillation, this should be quite sensitive to the bias voltage, since the minority-spin $\Delta_5$ band just touches but does not clearly cross the Fermi level. However, our experiments clarified that the period of the TMR oscillation hardly changes when applying the bias voltage up to $\pm\,0.5\,{\rm V}$ (see Fig. \ref{FigS3}), indicating that the minority-spin $\Delta_5$ state does not have a significant contribution to the TMR oscillation. All these facts allow us to neglect the minority-spin $\Delta_5$ state in our theory. Note also that the minority-spin $\Delta_2$ band crossing the Fermi level has another Fermi momentum $-k_2\,(>0)$ that is not considered in our theory. This is because we focus only on the states with a positive group velocity ($dE/dk_z>0$), based on our assumption that electrons tunnel from the left to right electrode.

\section{\label{Appe_E} Model for the interfacial spin-flip scattering}
It is expected that localized spins at interfacial Fe sites of the Fe/MgO/Fe(001) MTJ have spin fluctuations at finite temperature. Such spin fluctuations provide spin-flip scattering in the conduction electron states through the interfacial exchange interaction between localized spins at interfacial Fe sites and conduction electrons at the Fermi level. Based on the fact that only the majority-spin $\Delta_1$ and the minority-spin $\Delta_2$ bands cross $E_{\rm F}$ in the $\Delta$ line, the interfacial exchange interaction can be expressed by the following Hamiltonian:
\begin{eqnarray}
\nonumber H_{\rm ex}=\frac{J}{2}\,\sum_{{\bf k'}{\bf k}}\left[ S^{+}c^{\dagger}_{2{\bf k'}\downarrow}c_{1{\bf k}\uparrow}+S^{-}c^{\dagger}_{1{\bf k'}\uparrow}c_{2{\bf k}\downarrow} \right.\\
\left. +S^{z}(c^{\dagger}_{1{\bf k'}\uparrow}c_{1{\bf k}\uparrow}-c^{\dagger}_{2{\bf k'}\downarrow}c_{2{\bf k}\downarrow})\right],\label{Hex}
\end{eqnarray}
where $S^{\pm}=S^x \pm iS^y$ and $S^z$ are the spin operators for the localized spin at an interfacial Fe site, $c^{\dagger}_{1{\bf k}\uparrow}$ ($c^{\dagger}_{2{\bf k}\downarrow}$) is the creation operator of an electron with wave vector ${\bf k}$ in the majority-spin $\Delta_1$ (minority-spin $\Delta_2$) band, and $J\,(>0)$ is the coupling constant. This kind of exchange interaction between localized spins and conduction electrons in Bloch states was derived by Schrieffer and Wolff for the first time \cite{1966Schrieffer-PR}. Equation (\ref{Hex}) is a natural extension of the expression by Schrieffer and Wolff to the case near the interface of the Fe/MgO/Fe(001) MTJ. We assume that the localized spin has $S=2$ because of six $3d$ valence electrons in Fe. If we define the one-body energy for the majority-spin $\Delta_1$ (minority-spin $\Delta_2$) band as $\epsilon_{1,{\bf k}}$ ($\epsilon_{2,{\bf k}}$), the total Hamiltonian at the Fermi level is given by the following $2 \times 2$ matrix in the subspace spanned by the states $|S_z=1,k_1 \uparrow \rangle \equiv |k_1 \uparrow \rangle$ and $|S_z=2,k_2 \downarrow \rangle \equiv |k_2 \downarrow \rangle$:
\begin{equation}
H=\left(
\begin{array}{cc}
\bar{\epsilon}_{1,k_1} & J \\
J & \bar{\epsilon}_{2,k_2} \\
\end{array}
\right),\label{Hmat}
\end{equation}
where $\bar{\epsilon}_{1,k_1}=\epsilon_{1,k_1}+J/2$ and $\bar{\epsilon}_{2,k_2}=\epsilon_{2,k_2}-J$.

This real symmetric matrix $H$ can be diagonalized as
\begin{equation}
U^T H\, U=\left(
\begin{array}{cc}
E_+ & 0 \\
0 & E_- \\
\end{array}
\right)
\end{equation}
by using an orthogonal matrix
\begin{equation}
U=\left(
\begin{array}{cc}
u & -v \\
v & u \\
\end{array}
\right)\label{Umat}
\end{equation}
with $u$ and $v$ satisfying the relation $u^2+v^2=1$. Here, $E_{\pm}$, $u^2$, and $v^2$ are given by
\begin{eqnarray}
E_{\pm}=\frac{\bar{\epsilon}_{1,k_1}+\bar{\epsilon}_{2,k_2}}{2} \pm \omega \label{eival}\\
u^2=\frac{1}{2}\left( 1+ \frac{\bar{\epsilon}_{1,k_1}-\bar{\epsilon}_{2,k_2}}{2\,\omega} \right),\\ 
v^2=\frac{1}{2}\left( 1- \frac{\bar{\epsilon}_{1,k_1}-\bar{\epsilon}_{2,k_2}}{2\,\omega} \right),\label{u2_v2}
\end{eqnarray}
with
\begin{equation}
\omega=\sqrt{\left( \frac{\bar{\epsilon}_{1,k_1}-\bar{\epsilon}_{2,k_2}}{2} \right)^2+J^2}.
\end{equation}
The eigenstates
\begin{eqnarray}
|+\rangle &&\equiv \left(
\begin{array}{c}
u \\
v \\
\end{array}
\right)=u\,|k_1 \uparrow \rangle + v\,|k_2 \downarrow \rangle,\label{+state}\\
|-\rangle &&\equiv \left(
\begin{array}{c}
-v \\
u \\
\end{array}
\right)=-v\,|k_1 \uparrow \rangle + u\,|k_2 \downarrow \rangle\label{-state}
\end{eqnarray}
satisfy the eigenvalue equation
\begin{equation}
H | \pm \rangle = E_\pm | \pm \rangle.
\end{equation}
If we put $\epsilon_{1,k_1}=\epsilon_{2,k_2}=E_{\rm F}$ in Eqs. (\ref{eival})--(\ref{u2_v2}), we obtain $E_+=E_{\rm F}+J$, $E_-=E_{\rm F}-3J/2$, $u^2=4/5$, and $v^2=1/5$. Since $J$ is usually sufficiently smaller than $E_{\rm F}$, the eigenstates $|+\rangle$ and $|-\rangle$ are energetically close to $E_{\rm F}$ and thus can be considered as a transmission state for the tunneling problem \cite{short_note}. When $|k_1 \uparrow \rangle$ is the incident state in the left electrode, the state $|+\rangle$ ($|-\rangle$) with the dominant contribution from $|k_1 \uparrow \rangle$ ($|k_2 \downarrow \rangle$) corresponds to the transmitted wave in the right electrode for the parallel (antiparallel) magnetization state of the electrodes.

In this Appendix, we adopt a real symmetric Hamiltonian [Eq. (\ref{Hmat})] as an example. However, other Hamiltonians with different forms of spin-flip interactions may also be considered for the superposition of the majority-spin $\Delta_1$ and minority-spin $\Delta_2$ states. In general, the Hamiltonian is an Hermite matrix and can be diagonalized by the unitary matrix
\begin{equation}
U=\left(
\begin{array}{cc}
u & -v^\ast \\
v & u \\
\end{array}
\right),
\end{equation}
with $u$ and $v$ satisfying the relation $|u|^2+|v|^2=1$. This is the natural generalization of Eq. (\ref{Umat}). Similarly to Eqs. (\ref{+state}) and (\ref{-state}), the eigenstates $|+\rangle$ and $|-\rangle$ of the Hamiltonian are expressed as
\begin{eqnarray}
|+\rangle &&= \left(
\begin{array}{c}
u \\
v \\
\end{array}
\right)=u\,|k_1 \uparrow \rangle + v\,|k_2 \downarrow \rangle,\label{+state_2}\\
|-\rangle &&= \left(
\begin{array}{c}
-v^\ast \\
u \\
\end{array}
\right)=-v^\ast \,|k_1 \uparrow \rangle + u\,|k_2 \downarrow \rangle.\label{-state_2}
\end{eqnarray}
In the main manuscript, we use these states $|+\rangle$ and $|-\rangle$ for the transmitted waves in the parallel and antiparallel magnetization states, respectively. By multiplying the bra vector $\langle z|$ to Eqs. (\ref{+state_2}) and (\ref{-state_2}) and remembering that $|k_1 \uparrow \rangle$ and $|k_2 \downarrow \rangle$ are the eigenstates of the momentum operator, we obtain
\begin{eqnarray}
\psi_+(z)&\equiv&\langle z|+\rangle = u\,e^{ik_1z}+v\,e^{ik_2z},\\
\psi_-(z)&\equiv&\langle z|-\rangle = -v^\ast\, e^{ik_1z}+u\,e^{ik_2z},
\end{eqnarray}
which are the wave functions used for the transmitted waves in the main manuscript.

\section{\label{Appe_F} Analogy with the double-slit experiment}
By considering the superposition of the majority-spin $\Delta_1$ and the minority-spin $\Delta_2$ states for the transmitted wave function, the following factor appears in some terms in the denominator of the transmittance:
\begin{equation}
(a\,u\,e^{ik_1d}+b\,v\,e^{ik_2d})(a^{\ast}u^{\ast}e^{-ik_1d}+b^{\ast}v^{\ast}e^{-ik_2d}),\label{key-factor}
\end{equation}
where the definition of $u$, $v$, $k_1$, and $k_2$ are the same as those in the main manuscript and $a$ and $b$ are the coefficients determined by the boundary conditions. The cross terms of Eq. (\ref{key-factor}) are
\begin{equation}
a\,b^{\ast}u\,v^{\ast}e^{i(k_1-k_2)d}+a^{\ast}b\,u^{\ast}v\,e^{-i(k_1-k_2)d},
\end{equation}
which provide $\cos{((k_1-k_2)\,d)}$ and $\sin{((k_1-k_2)\,d)}$, leading to oscillations of the transmittances and the TMR ratio. Importantly, this is mathematically similar to the double-slit experiment in elementary quantum mechanics. Using the wave functions $\Psi_{\rm A}$ and $\Psi_{\rm B}$ passing through the windows A and B, the probability of finding an electron in the screen behind the windows is given by
\begin{equation}
|\Psi_{\rm A}+\Psi_{\rm B}|^2=(\Psi_{\rm A}+\Psi_{\rm B})(\Psi^{\ast}_{\rm A}+\Psi^{\ast}_{\rm B}),
\end{equation}
where the cross terms $\Psi_{\rm A}\Psi^{\ast}_{\rm B}+\Psi_{\rm B}\Psi^{\ast}_{\rm A}$ express the interference of the wave functions $\Psi_{\rm A}$ and $\Psi_{\rm B}$. This analogy provides us an intuitive physical picture that the TMR oscillation is driven by the interference of the majority-spin $\Delta_1$ and the minority-spin $\Delta_2$ wave functions in the transmitted wave, as shown schematically in Fig. 1(d).

\section{\label{Appe_G} Experimental method}
\begin{figure}
\includegraphics[width=5.5cm]{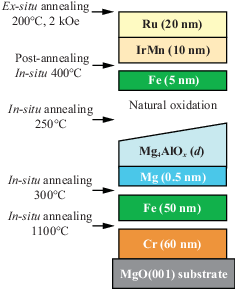}
\caption{\label{FigS2} Schematic MTJ stacking structure.}
\end{figure}
We fabricated an MTJ structure at room temperature using magnetron sputtering and electron-beam (EB) evaporation at a base pressure of $< 1 \times 10^{-6}\,{\rm Pa}$ on a single crystal MgO(001) substrate. The stack structure is MgO substrate/Cr (60)/Fe (50)/Mg (0.5)/wedge-shaped Mg$_4$AlO$_x$ (MAO) ($d=$1.0--3.0)/natural oxidation/Fe (5)/Ir$_{22}$Mn$_{78}$ (10)/Ru (20) (thickness in nm) (Fig. \ref{FigS2}). The MAO barrier was deposited using EB evaporation of a sintered MAO block with nominal Mg/Al atomic ratio\,=\,4. Because of such a high Mg concentration, the present MAO has a similar electronic structure as MgO considered in our theory. Back-pressure and deposition rates were $\sim 4 \times 10^{-6}\,{\rm Pa}$ and $8 \times 10^{-3}\,{\rm nm/s}$, respectively. The wedge-shaped MAO layer was prepared using a linear motion shutter. The Cr buffer layer was post-annealed at 1100$^\circ$C for 10 min using an infrared rapid thermal annealing system. The bottom  and top Fe layers were annealed at 300$^\circ$C and 400$^\circ$C, respectively. The barrier was post-annealed at 250$^\circ$C followed by 300\,s natural oxidation after cooling to room temperature. After the deposition, the unpatterned films were annealed in a magnetic field at 200$^\circ$C along the MgO[110] $\parallel$ Fe[100] direction. Photolithography and Ar-ion etching were used to pattern the stacks into 39 $\mu{\rm m}^2$ area elliptical junctions with a long axis parallel to the Fe[100] easy axis. The magnetotransport properties were characterized by a DC 4-probe method using a sourcemeter and a nanovoltmeter. The TMR ratio is defined as $(R_{\rm AP}-R_{\rm P})/R_{\rm P}$, where $R_{\rm P}$ ($R_{\rm AP}$) is the resistance in the parallel (antiparallel) magnetization state.

\section{\label{Appe_H} Further experimental results}
\begin{figure}
\includegraphics[width=7.0cm]{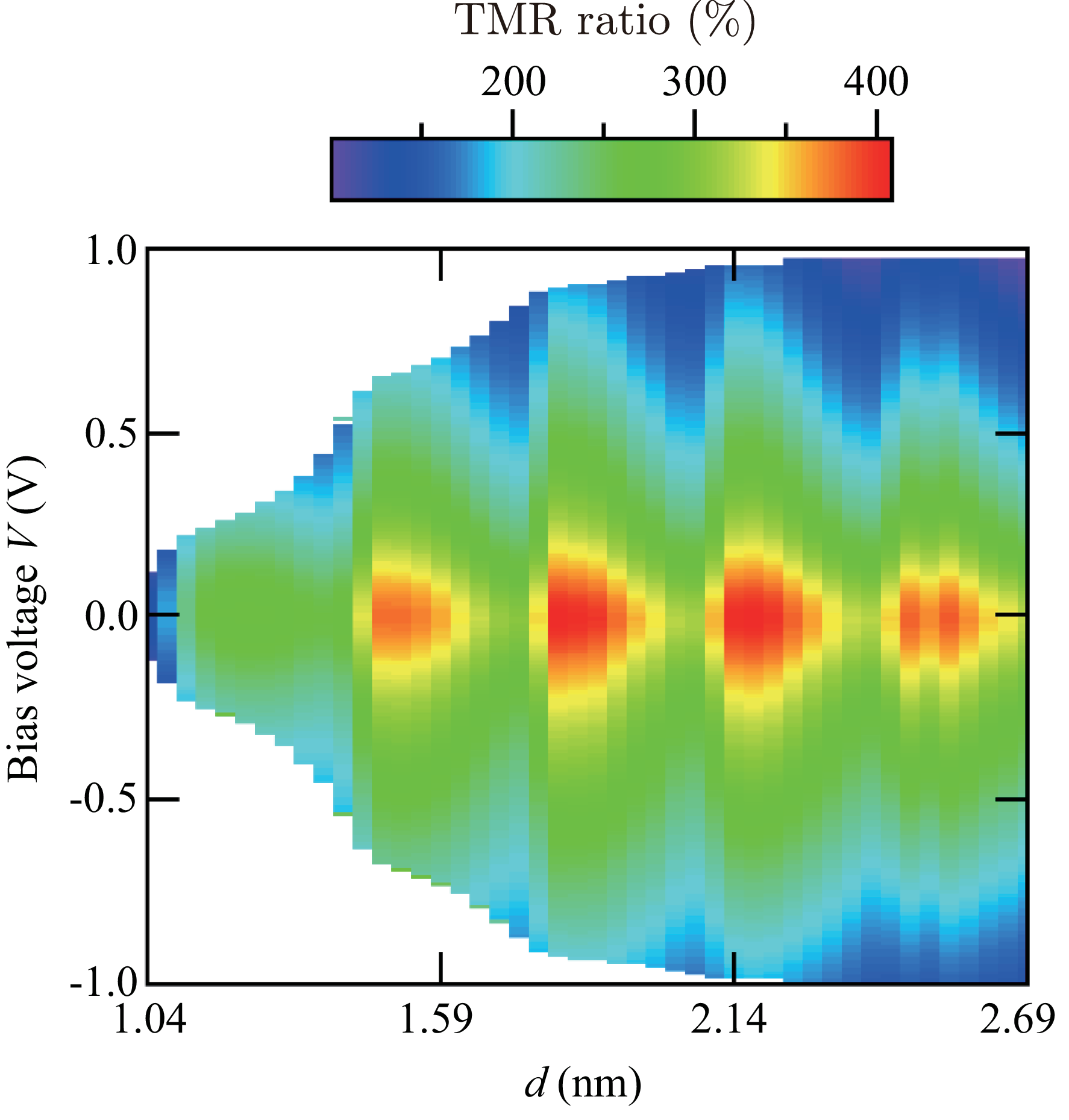}
\caption{\label{FigS3} Two-dimensional map of the TMR ratio in Fe/MAO/Fe(001) as functions of the barrier thickness $d$ and the bias voltage $V$.}
\end{figure}
\begin{figure}
\includegraphics[width=8.5cm]{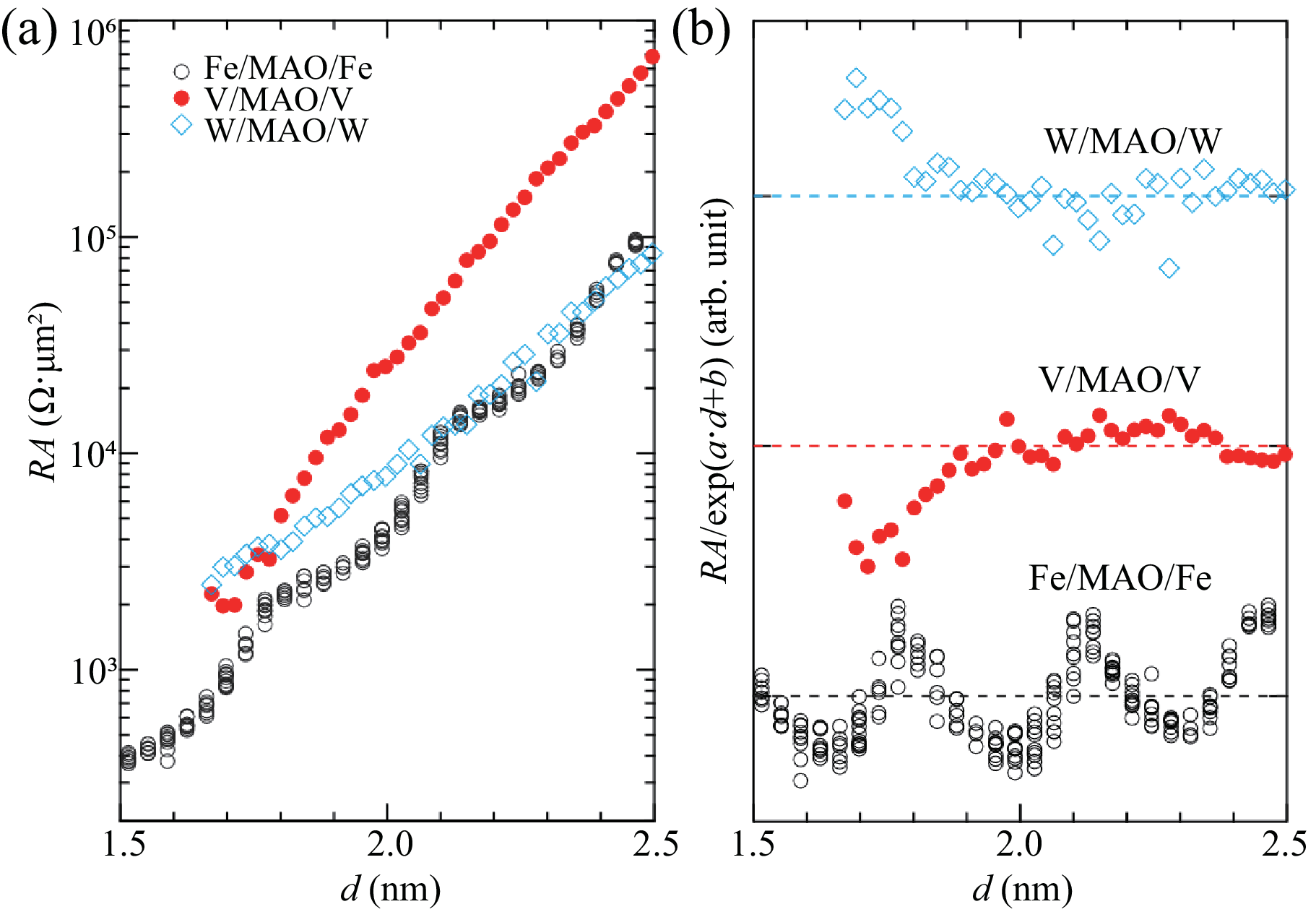}
\caption{\label{FigS4} (a) Barrier thickness dependences of $RA$ for Fe/MAO/Fe(001) in the parallel magnetization state, V/MAO/V(001), and W/MAO/W(001) at room temperature. (b) The same as (a) but are divided by the exponentially increasing factor $\exp{(a \cdot d+b)}$. The dashed lines in (b) are the baselines of each plot.}
\end{figure}
As shown in Fig. \ref{FigS3}, we applied bias voltages up to $\sim \pm\, 0.5\,{\rm V}$ to the Fe/MAO/Fe(001) MTJ, which clarified that the period of the TMR oscillation hardly changes under the bias voltages. The TMR effect under bias voltages can be discussed by introducing an energy shift in the band structure between the left and right electrodes. Namely, when electrons tunnel from the left to right electrode under a bias voltage $V\,(>0)$, we shift the band structure in the left electrode by $+{\rm e}V$ relative to the band structure in the right electrode. Then, we discuss the electron tunneling from occupied states in the left electrode to unoccupied states in the right electrode, which occurs in the energy window $E_{\rm F}<E<E_{\rm F}+{\rm e}V$ with $E_{\rm F}$ being the original Fermi energy. Let us here focus on the unoccupied states with $E_{\rm F}<E<E_{\rm F}+{\rm e}V$ in the right electrode, since the superposition of wave functions in this electrode is the key to the resistance oscillation. When we examine the band structures of Fe changing the energy $E$ from $E_{\rm F}$ to $E_{\rm F}+{\rm e}V$ (${\rm e}V=0.5\,{\rm eV}$), the absolute value of $k_1$ ($k_2$) increases (decreases) as seen from Figs. 1(b) and 1(c) (note that $k_2<0$), and thus the value of $k_1 - k_2$ is almost unchanged. This indicates that the period of the TMR oscillation $2\pi/(k_1-k_2)$ in our theory hardly changes under the bias voltages, consistent with the experimental observation. The case of a negative bias voltage $V\,(<0)$ can be discussed in a similar way. These facts can justify our idea of considering the superposition of the majority-spin $\Delta_1$ and the minority-spin $\Delta_2$ wave functions.

We also conducted experiments using non-magnetic bcc transition metals (V and W) as electrodes of tunnel junctions. The stacking structures of these junctions are similar to that of the Fe/MAO/Fe(001) MTJ except that Fe is replaced by V or W, i.e., MgO substrate/Cr (60)/{\it X} (50)/Mg (0.5)/wedge-shaped MAO ($d=$1.0--3.0)/natural oxidation/{\it X} (5)/Ir$_{22}$Mn$_{78}$ (10)/Ru (20) (thickness in nm), where {\it X} is V or W. In these junctions, the Cr buffer layer was post-annealed at 600$^\circ$C for 45 min using a lamp heater system. As shown in Figs. \ref{FigS4}(a) and \ref{FigS4}(b), we found no oscillations in $RA$ values in both V/MAO/V(001) and W/MAO/W(001), indicating that the spin-polarized band structure is essential for the occurrence of the resistance oscillation. These results also support our theory, since the superposition of the majority- and minority-spin wave functions due to the spin-flip scattering is the key to our theory.

\section{\label{Appe_I} Relation with experiments using Heusler alloys}
\begin{figure}
\includegraphics[width=8.5cm]{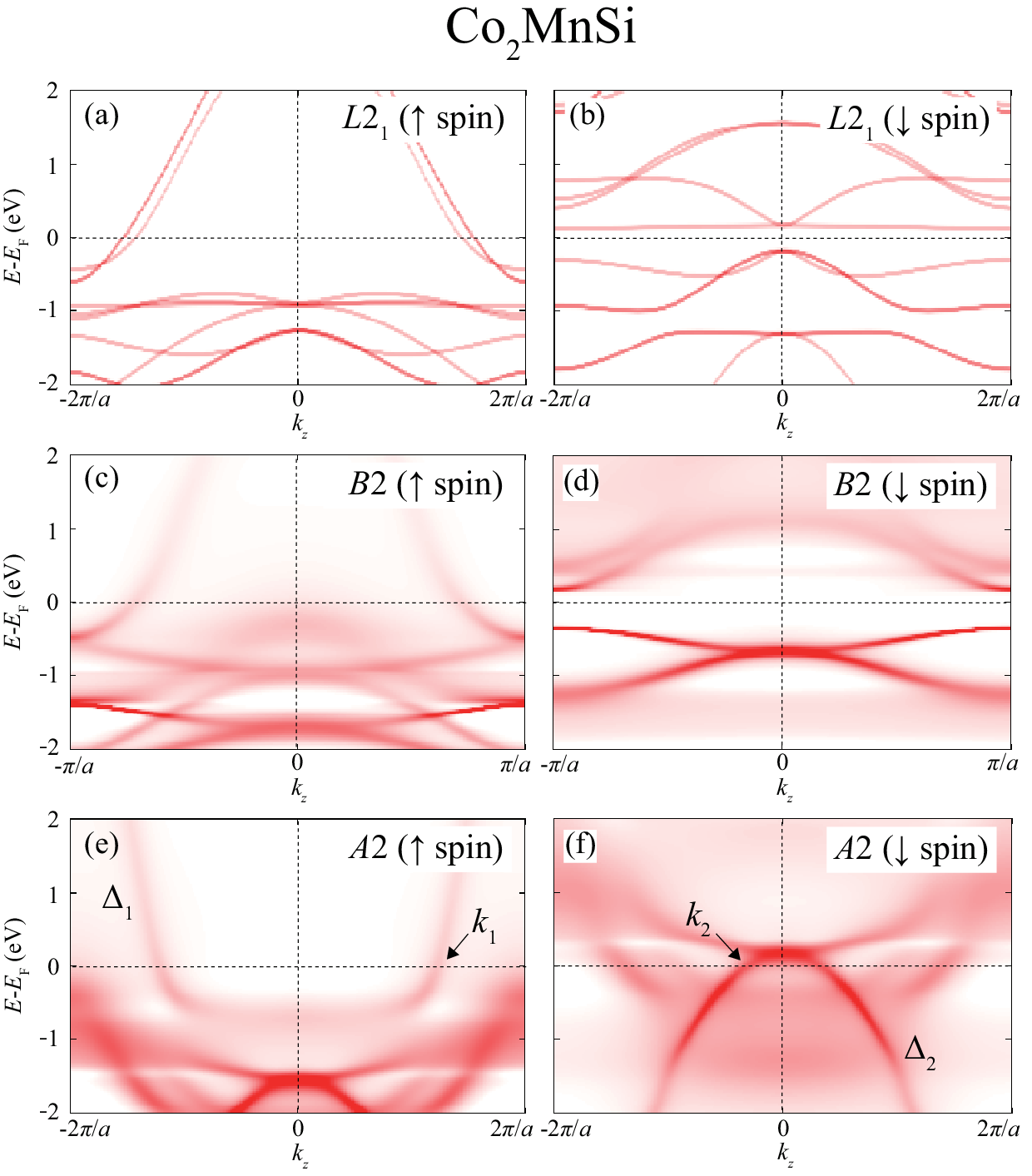}
\caption{\label{FigS5} Band structures of Co$_2$MnSi in the (a,b) $L2_1$, (c,d) $B2$, and (e,f) $A2$ orders along the $\Delta$ line. Panels (a,c,e) are for the majority-spin state and panels (b,d,f) are for the minority-spin state.}
\end{figure}
\begin{figure}
\includegraphics[width=8.5cm]{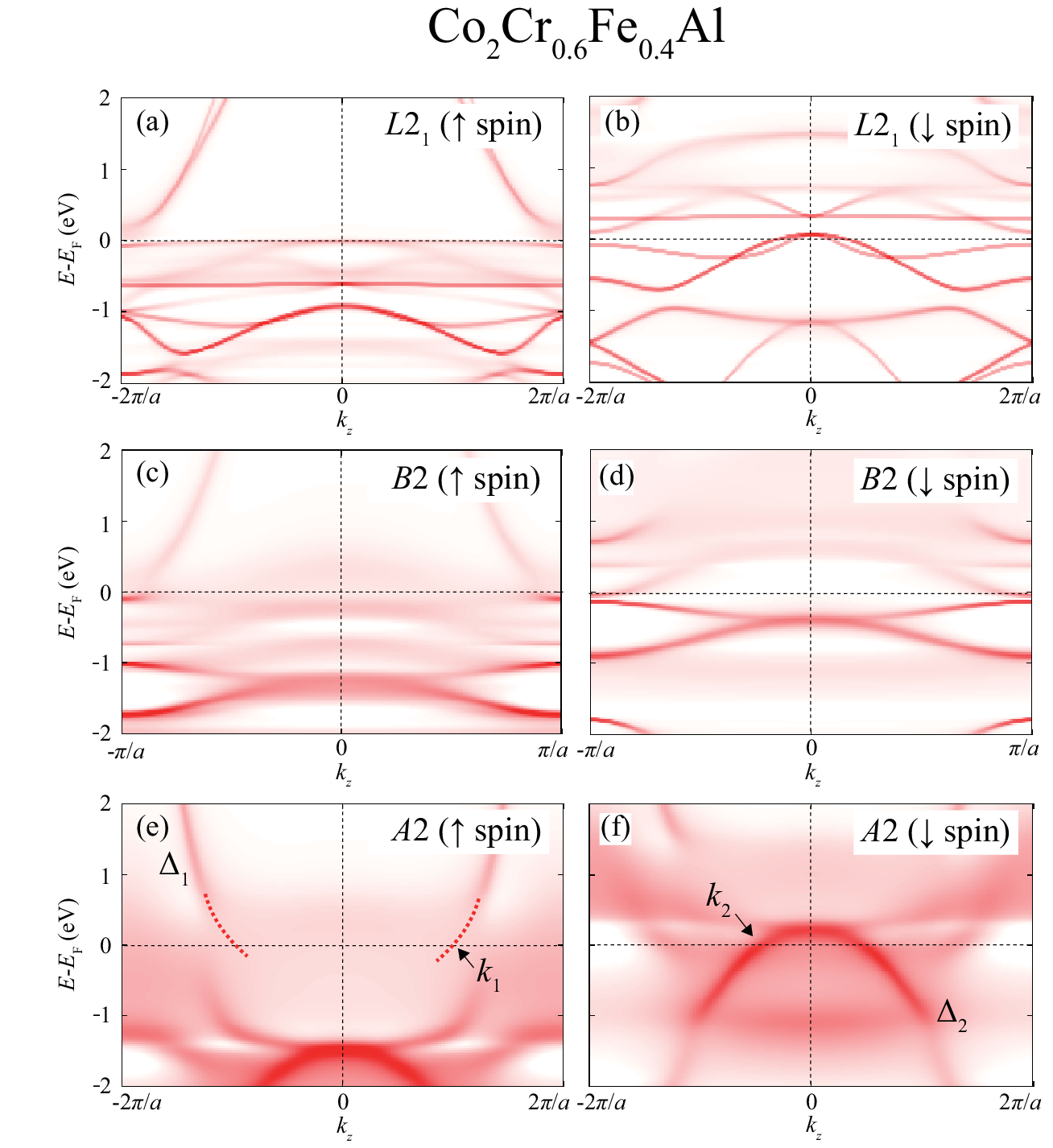}
\caption{\label{FigS6} The same as Fig. \ref{FigS5} but for Co$_2$Cr$_{0.6}$Fe$_{0.4}$Al.}
\end{figure}
Some previous experimental studies have reported the occurrence of the TMR oscillation in MTJs with Heusler alloys (Co$_2$MnSi and Co$_2$Cr$_{0.6}$Fe$_{0.4}$Al) \cite{2008Ishikawa-JAP,2010Marukame-PRB}. We here discuss the relation between these experiments and our theory by calculating band structures of the Heusler alloys. We have employed the Korringa-Kohn-Rostoker (KKR) method \cite{1947Korringa-Physica,1954Kohn-PR} based on the density-functional theory, implemented in the AkaiKKR software package \cite{Akai-KKR}. Here, the generalized gradient approximation \cite{1996Perdew-PRL} was used for the exchange-correlation energy, and disordered states of the Heusler alloys were treated within the coherent potential approximation \cite{1967Soven-PR}. We used the following lattice constants for Co$_2$MnSi (Co$_2$Cr$_{0.6}$Fe$_{0.4}$Al) in our calculations: $a=5.645\,(5.824)\,{\rm \AA}$ for the $L2_1$ order and $a=2.822\,(2.912)\,{\rm \AA}$ for the $B2$ and $A2$ orders.

Figures \ref{FigS5}(a) and \ref{FigS5}(b) show the majority- and minority-spin band structures of Co$_2$MnSi with the ideal $L2_1$ order calculated along the $\Delta$ line ($k_z$ line at $k_x=k_y=0$). Here, the $L2_1$ order denotes a perfectly ordered crystal structure composed of four interpenetrating fcc lattices, two of which are occupied by Co atoms and another two by Mn and Si atoms, respectively. We see that some bands cross $E_{\rm F}$ in the majority-spin state [Fig. \ref{FigS5}(a)], while no band crosses $E_{\rm F}$ in the minority-spin state [Fig. \ref{FigS5}(b)], consistent with the well-known fact that Co$_2$MnSi with the $L2_1$ order has a half-metallic band structure \cite{1995Ishida-JPSJ,2002Galanakis-PRB}. In this case, the present theory predicts the absence of the TMR oscillation due to the absence of the superposition between the majority- and minority-spin states at $E_{\rm F}$. However, Heusler alloys in thin-film systems such as MTJs are not free from the effect of atomic disorder; previous studies have found that the degree of the $L2_1$ order is usually much smaller than 1, indicating the existence of atomic disorder from the $L2_1$ order \cite{2008Gaier-JAP}. We here consider two typical disordered states called the $B2$ and $A2$ orders in Co$_2$MnSi: Mn and Si atoms are randomly disordered in the $B2$ order and all the atoms are disordered in the $A2$ order. In Figs. \ref{FigS5}(c) and \ref{FigS5}(d), we show the majority- and minority-spin band structures in the $B2$ order. Since no minority-spin band crosses $E_{\rm F}$, we cannot expect the existence of the TMR oscillation from our theory similarly to the case of the $L2_1$ order. In contrast, in the $A2$ order, the majority-spin $\Delta_1$ and the minority-spin $\Delta_2$ bands cross $E_{\rm F}$ as shown in Figs. \ref{FigS5}(e) and \ref{FigS5}(f), indicating the existence of the TMR oscillation. Based on our theory, the period of the oscillation is calculated as $2\pi/(k_1-k_2) \approx 3.4\,{\rm \AA}$, which is a bit larger than $2.8\,{\rm \AA}$ obtained experimentally \cite{2008Ishikawa-JAP}. Note here that this difference is within the acceptable range since the amplitude of the TMR oscillation in this experiment is very small and the experimental period includes unavoidable observational errors.

In Figs. \ref{FigS6}(a)--\ref{FigS6}(f), we show majority- and minority-spin band structures of Co$_2$Cr$_{0.6}$Fe$_{0.4}$Al calculated in the $L2_1$, $B2$, and $A2$ orders. In the $L2_1$ and $B2$ orders, it is difficult to predict from our theory whether the TMR oscillation can occur because of complex band structures with flat dispersions near $E_{\rm F}$. In contrast, in the $A2$ order, the qualitative features of band structures are similar to those in the $A2$ order in Co$_2$MnSi; the majority-spin $\Delta_1$ and the minority-spin $\Delta_2$ bands with large dispersions cross $E_{\rm F}$ as shown in Figs. \ref{FigS6}(e) and \ref{FigS6}(f). Although the majority-spin $\Delta_1$ band is smeared out near $E_{\rm F}$ due to the effect of the disorder, the period of the oscillation can be roughly estimated as $2\pi/(k_1-k_2) \approx 3.5\,{\rm \AA}$, which is close to $3.2\,{\rm \AA}$ obtained experimentally \cite{2010Marukame-PRB}. All these comparisons suggest that there is no inconsistency between our findings and previous experiments using Heusler alloys, supporting our theory for the TMR oscillation.


\begin{thebibliography}{99}
\bibitem{remark_superposition} We also considered a similar superposition of wave functions in the incident and reflection waves. This additional analysis clarified that these do not affect the oscillatory behavior of the transmittance given by the superposition in the transmitted wave. Thus, we omit these effects for simplicity.
\bibitem{remark_AP-state} This is because $T_{{\rm AP},\uparrow}$ can be calculated by using $\psi_{\rm R}(z)=C\,(-v^{\ast}\,e^{ik_1 z}+u\,e^{ik_2 z})$ instead of Eq. (\ref{new_tun_prob}).
\bibitem{remark_osci-phase1} In our results, differences in both the shape and phase between the $T^{-1}_{{\rm P},\uparrow}$ and $T^{-1}_{{\rm AP},\uparrow}$ oscillations provide the TMR oscillation. Although the oscillation shapes of $T^{-1}_{{\rm P},\uparrow}$ and $T^{-1}_{{\rm AP},\uparrow}$ look quite similar, these have a slight difference owing to the difference in the effective mass and the replacement of the coefficients ($u$ and $v$) in the expression of the transmittance.
\bibitem{remark1} TMR ratios at low temperature are more than twice as high as those at room temperature and can be reproduced by using a smaller value of $\alpha$.
\bibitem{remark_osci-amp} The amplitude of the $RA$ oscillation hardly changes as the barrier thickness $d$ increases, consistent with the results in Fig. \ref{Fig2}. In our theory, the amplitude of the transmittance oscillation is determined by the values of $u$ and $v$ in Eq. (\ref{new_tun_prob}). Since $u$ and $v$ are assumed to be constant, the amplitude of the oscillation hardly changes with increasing $d$. This assumption is reasonable, since the interfacial exchange interaction determining $u$ and $v$ is independent on $d$.
\bibitem{remark2} We introduced non-oscillatory terms $\beta\, e^{-2 \kappa d}$ decreasing exponentially as the barrier thickness increases.
\bibitem{remark_d-shift} For a better comparison with experimental results, we applied a shift in $d$ by 0.8\,{\AA} for Eq. (\ref{TMR2}).
\bibitem{short_note} Since $E_{+}$ and $E_{-}$ are not exactly equal to $E_{\rm F}$, one may think that $|+\rangle$ and $|-\rangle$ cannot be regarded as the transmission states in a rigorous sense. We here show that both $E_{+}$ and $E_{-}$ can be set exactly equal to $E_{\rm F}$ by considering a tiny shift $\Delta k_{i}$ in the Fermi momentum $k_i$ ($i=1,2$). For the parallel magnetization state, we choose $\Delta k_{1}$ such that $\epsilon_{1,k_1+\Delta k_1}=E_{\rm F}-3J/2$ and diagonalize the matrix obtained by replacing $\bar{\epsilon}_{1,k_1}$ by $\bar{\epsilon}_{1,k_1+\Delta k_1}$ in Eq. (\ref{Hmat}), leading to $E_{+}=E_{\rm F}$. For the antiparallel magnetization state, we can obtain $E_{-}=E_{\rm F}$ by choosing $\Delta k_{1}$ and $\Delta k_{2}$ such that $\epsilon_{1,k_1+\Delta k_1}=E_{\rm F}+J/2$ and $\epsilon_{2,k_2+\Delta k_2}=E_{\rm F}+2J$. Note here that the diagonalization for $E_{+}$ and that for $E_{-}$ are independent calculations, since these treat different (i.e., parallel and antiparallel) magnetization states of the electrodes. The states $|+\rangle$ and $|-\rangle$ obtained here are qualitatively similar to those obtained without introducing $\Delta k_{i}$ [Eqs. (\ref{+state}) and (\ref{-state})], since $|\Delta k_{i}| \ll |k_{i}|$ owing to $J \ll E_{\rm F}$. We can always choose such values of $\Delta k_{i}$, because $\epsilon_{i,k}$ ($i=1,2$) is continuous spectrum and the fact that $J \ll E_{\rm F}$.
\end{thebibliography}

\end{document}